# Particle-in-cell simulation of x-ray wakefield acceleration and betatron radiation in nanotubes

Xiaomei Zhang,[1,2] Toshiki Tajima,[2] Deano Farinella,[2] Youngmin Shin,[3] Gerard Mourou,[4] Jonathan Wheeler,[4] Peter Taborek,[2] Pisin Chen,[5] Franklin Dollar,[2] and Baifei Shen[1]

[1]*Shanghai Institute of Optics and Fine Mechanics, Chinese Academy of Sciences, Shanghai 201800, China*
[2]*Department of Physics and Astronomy, UC, Irvine, California 92697, USA*
[3]*Northern Illinois University and Fermi National Accelerator Laboratory, Dekalb, Illinois 60115, USA*
[4]*DER-IZEST, École Polytechnique, 91128 Palaiseau Cedex, France*
[5]*Department of Physics & Leung Center for Cosmology and Particle Astrophysics, National Taiwan University, Taipei 10617, Taiwan*
(Received 26 January 2016; published 18 October 2016)

Though wakefield acceleration in crystal channels has been previously proposed, x-ray wakefield acceleration has only recently become a realistic possibility since the invention of the single-cycled optical laser compression technique. We investigate the acceleration due to a wakefield induced by a coherent, ultrashort x-ray pulse guided by a nanoscale channel inside a solid material. By two-dimensional particle-in-cell computer simulations, we show that an acceleration gradient of TeV/cm is attainable. This is about 3 orders of magnitude stronger than that of the conventional plasma-based wakefield accelerations, which implies the possibility of an extremely compact scheme to attain ultrahigh energies. In addition to particle acceleration, this scheme can also induce the emission of high energy photons at $\sim O(10\text{--}100)$ MeV. Our simulations confirm such high energy photon emissions, which is in contrast with that induced by the optical laser driven wakefield scheme. In addition to this, the significantly improved emittance of the energetic electrons has been discussed.

DOI: 10.1103/PhysRevAccelBeams.19.101004

## I. INTRODUCTION

Electrons can be accelerated to high energies in the wakefield formed when a short pulse laser or beam passes through a plasma [1,2]. Experiments have shown that the GeV energy gain can be obtained over centimeter-scale distances within a gaseous plasma by riding on the wakefield excited by optical lasers [3–7]. Laser wakefield theory [1,8,9] shows that for a given laser, the energy gain and accelerating length are both inversely proportional to the plasma density. This means that the lower the gas density, the longer the acceleration distance required to reach greater energies, an undesirable condition for achieving the goal of ultrahigh energies. Motivated by such considerations, utilization of metallic crystals was proposed in the 1980s [10–16], where TeV/cm acceleration gradient was anticipated. This includes the cases of wakefield acceleration in metallic crystal channels. Another advantage of solid-state guided acceleration is that such a system can naturally provide the mechanism radiation damping. Here the accelerated particle beam emittance, i.e., the transverse momentum, can be dramatically damped through channeling radiation to the ground state of the channels [17]. Under such a scenario, one may even envision head-on collisions of ultrahigh energy particles inside these microscopic channels at their ground states, where the classical concept of luminosity is in the so-called *quantum luminosity* regime, which promises a much higher collision rate [18]. On the other hand, a disadvantage of metallic channels is its high collision frequency with the metallic electrons [19]. This may be alleviated by adopting nanoholes [14,20–22].

One of the most important motivating factors of the present paper, in addition to the above, however, is the recent advent of the breakthrough in the laser compression technique that could open a door for an evolution of any possibility of a coherent intense x-ray laser pulse in attosecond regimes. The recently proposed scheme of ultrashort, coherent x-ray pulse generation derived from the new optical laser compression [23] into a single-cycled optical pulse, in combination with the relativistic surface compression [24] of such an optical laser into an x-ray laser pulse, provides an attractive possibility to realize such an ultrahigh acceleration gradient, for a compact solid-state accelerator scheme to accelerate particles to ultrahigh energies. The thin film compression [23] can be a simple elegant method of an ultrafast intense optical laser into a single-cycled optical laser pulse with high efficiency (such as ~90%). In turn, such a single-cycled optical laser pulse may be relativistically compressed by the well-known relativistic surface compression [24] into a single-cycled x-ray laser pulse, whose photon energies may be up to ∼10 keV [24]. In principle the frequency of the driving x rays can match the much higher critical density $n_c$, provided by the conductive solid material which depends inversely to the square of the laser wavelength:

$$n_c(\lambda_L) = \frac{\pi m_e c^2}{e^2 \lambda_L^2} \approx (1.1 \times 10^{21} \ [\text{cm}^{-3}])/\lambda_L^2 [\mu\text{m}].$$

Here $c$ is the light speed in vacuum, $m_e$ is the electron mass, $e$ is the electron charge and $\lambda_L$ is the laser wavelength. This high critical density by the x-ray laser allows us the additional advantage, i.e., the long dephasing length [1,20]. One point to notice here is the following. Unlike optically photons, photons of x-ray regimes can see even shallowly bound electrons whose binding energy is less than the photon energy (such as 10 keV). Thus, even if the material is, for the usual purpose of condensed materials, not a plasma but a bound-state condensed material, x-ray photons see these shallowly bound electrons as if they are (efficiency) free electrons. We, therefore, treat such electrons as free electrons as in a plasma. In the following when we call plasma in the solid density for x rays, we mean such electrons in the condensed material. Additional comments as to the use of the "collisionless" model of the particle-in-cell (PIC) simulation are below and done here. For one, the time scale of the intense x-ray driven electron dynamics is on the order of attoseconds (or even zeptoseconds), so that collisional effects may be ignored in these short time scales. Second, in terms of a longer time scale dynamics, we introduce the nanotube materials so that accelerated electron dynamics in the nanotube remains to be collisionless over an extended time scale of propagation dynamics. For these two reasons of the first order importance is the collisionless dynamics of what we call the plasma driven by intense x ray. However, unlike in gaseous plasma driven by an optical laser, it is anticipated that the quantum mechanical radiation processes could be far more important. That is why we incorporate the quantum radiative effects in our study.

Functional nanomaterials such as carbon nanotubes have a large degree of dimensional flexibility and allow for a greater than 10 TV/m acceleration gradient. Accordingly, compact structures to obtain ultrahigh energy gain can in principle be realized through the state-of-the-art nanotube technology [20]. A plasma channel is useful for guiding the laser [25,26], especially for a small laser spot. For an x-ray beam with a spot size at the nanometer level, stable propagation is important for the purpose of wakefield generation and acceleration. Available nanometer structures [27–30] such as porous alumina as shown in Fig. 1 [28,30] and carbon nanotubes provide an excellent prospect to guide the x-ray pulse while additionally guiding and collimating the high energy beam being accelerated, providing well-organized beam optics control. In such a material while the nanohole provides a good collisionless particle propagation, the surrounding nanomaterial supports the robust wakefield, where the x-ray laser aperture may cover a sufficient area. Furthermore, the honeycomb repeated structure allows us to easily raster the x-ray laser pulses with repeated high repetition irradiations. Also importantly, the beam emittance is damped through channeling, or betatron, radiation as mentioned above [18].

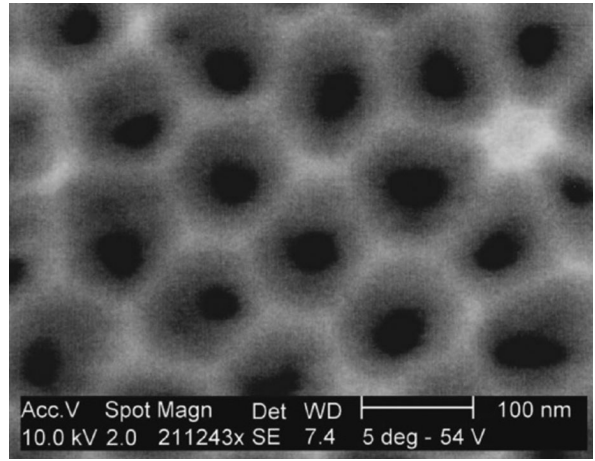

FIG. 1. SEM image of the top surface of a porous alumina sample [30].

Motivated by these points, in this paper we explore an x-ray wakefield accelerator within a nanotube. By comparing the two cases of a coherent, ultrafast x-ray pulse and 1 eV optical laser, the identical wake structure after normalization to the laser wavelength has been confirmed in the limit of the collisionless PIC modeling. However, the radiations in these two cases are quite different when the effect of quantum electrodynamics (QED) is considered. Photons with energy $\sim O(10 \ \text{MeV})$ have been generated in the x-ray laser case for the much stronger wakefield and the emittance of energetic electrons becomes 3 orders of magnitude lower. Based on the simulation results, we summarize the wakefield scalings with laser intensity, nanotube radius, and nanotube density.

## II. COMPARISON BETWEEN THE BASE CASES IN TERMS OF A NANOTUBE AND UNIFORM DENSITY

In the kind of intense irradiation of such an x-ray laser onto a solid material regardless if it is a metal or dielectric, the high energy of photons $[\varepsilon = \hbar\omega_L \sim O(\text{keV})$, where $\omega_L$ is the laser frequency] and the high intensity $[a_0 = eE_L/m_e\omega_L c = O(1)$, where $E_L$ is the laser field] both contribute to allow us to treat the material well approximated as plasma at a metallic density as mentioned in the Introduction [10–12,20]. The high energy of the photons makes a substantial amount of electrons [either the electrons in the conduction band or in the shallowly (<1 keV) bound electrons] respond to the x-ray fields directly. The high intensity of the x-ray pulse results in the instantaneous ionization of some of the bound electrons per atomic site, thereby contributing to free electrons. Even some remaining bound electrons may be treated a solid plasma as shown in Ref. [31], where additional optical phonon modes and Buchsbaum resonances are allowed.

TABLE I. Summary of the laser and plasma parameters for our base case.

| Laser wavelength $\lambda_L$ | Peak amplitude $a_0$ | Width radius $\sigma_L$ | Length radius $\sigma_x$ | Plasma density $n_{\text{tube}}$ | Tube radius $\sigma_{\text{tube}}$ |
|---|---|---|---|---|---|
| 1 nm | 4 | 5 nm | 3 nm | $5 \times 10^{24}$ W/cm$^3$ | 2.5 nm/0 nm |
| 1 $\mu$m | 4 | 5 $\mu$m | 3 $\mu$m | $5 \times 10^{18}$ W/cm$^3$ | 2.5 $\mu$m/0 $\mu$m |

Two-dimensional (2D) PIC simulations have been performed by using the EPOCH code written in SI units [32]. The simulation box is 60 nm(x) × 100 nm(y), which corresponds to a moving window with 3000 × 500 cells and ten particles per cell. For the base case, the laser and plasma parameters are listed in Table I. The laser pulse of wavelength $\lambda_L = 1$ nm (corresponding 1 keV x-ray laser), the peak normalized amplitude used is $a_0 = 4$, which means the pulse peak intensity is $2.2 \times 10^{25}$ W/cm$^2$. The tube wall density is given in terms of the critical density by $n_{\text{tube}} = 4.55 \times 10^{-3} n_c$. That is, for modeling the nanotube, a solid tube with wall density of $n_{\text{tube}} = 5 \times 10^{24}$/cm$^3$ is used. The tube location is $2\lambda_L < x < 8000\lambda_L$ and $-50\lambda_L < y < 50\lambda_L$. At $t = 0$, the laser pulse enters the simulation box from the left boundary.

Figures 2 and 3 show the comparison between the nanotube case and uniform density case driven by the x-ray pulse, in which a small spot size of $\sigma_L = 5$ nm over a short length of $\sigma_x = 3$ nm is chosen according to the approach proposed in Ref. [23]. For the uniform density case, see Figs. 2(b), 2(d), 2(f), 2(h) and 3, the Rayleigh length is short due to the small spot size, so the laser pulse quickly diverges as it propagates. Due to the defocusing laser field, the laser field decreases rapidly with the propagation distance, and thus the strong longitudinal wakefield only keeps a very short time and then goes weaker and gradually disappears. In this case, the driving pulse dissipated after propagating a distance of $2000\lambda_L$ and the wakefield is not stable during the whole process. However, in the nanotube case, as we see in Figs. 2(a), 2(c), 2(e), 2(g) and 3, the x-ray pulse maintains a small spot size that can be well controlled and guided by the surrounding nanotube walls. The induced wakefield stays stable and the short laser pulse continues propagating even after a distance of $4500\lambda_L$, which is more than twice that of the uniform density case. By comparison, we see the nanotube wakefield is akin to nonlinear wakefield in the bubble regime, while the uniform plasma wakefield seems to be a rather quasilinear wakefield. Both the longitudinal wakefield contributing to the accelerating force and transverse wakefield contributing to the focusing force on electrons are more stable and appropriate in the nanotube case. This stability over a long distance is important for the acceleration to obtain a high energy beam. Thus we see superior wakefield quality in a nanotube in comparison with the case in its comparable uniform medium.

To make a comparison, the simulations driven with an optical 1 eV laser pulse under analogous conditions are carried out. In this case when the same $a_0 = 4$ is used for the laser wavelength of $\lambda_L = 1$ $\mu$m, it corresponds to a laser peak intensity of $2.2 \times 10^{19}$ W/cm$^2$. It is expected that the wake structures are almost identical after all physical parameters are normalized by the laser wavelength and the simulation results confirm this. Considering the real physical parameters, it can be found the wakefield is higher than 2 TV/cm when driven by the x-ray pulse, which is 3 orders higher than that of the optical laser case. This means the energy gain gradient is 2 TeV/cm instead of 2 GeV/cm and opens the possibility to realize a very compact accelerator capable of reaching ultrahigh energies. In addition, the wakefield for the uniform plasma case can be estimated from $E_0 = a_0^{1\sim 2} m_e \omega_p c/e$, which is about $2.2 a_0^{1\sim 2}$ TV/cm using the parameters in the above simulations, where $\omega_p$ is the plasma frequency. This expected value agrees well with the observed one in Fig. 2, which means in the narrow limit of the tube, the wakefield scaling resembles that in the uniform plasma formulation.

Similar momenta (energy gains) are expected if the same ratio is kept between the laser and plasma wavelengths over one dephasing length, irrespective of the laser wavelength and background density. However, for the electron beam accelerated in the x-ray driven wakefield, one important signature is that the emittance can be improved significantly due to the much smaller size in the transverse dimension. As well known, beam emittance—related to both the transverse dimension and the electron momentum—is an important parameter with many applications requiring it to remain low. Similar energy gain is confirmed in Fig. 4, which shows the wakefield and the relativistic factor of the accelerated electrons driven by an x-ray pulse and an optical laser, respectively. The laser and plasma parameters are listed in Table II for the electron acceleration case. Here a higher $a_0 = 10$ is used to ensure the occurrence of self-injection. Figure 5 demonstrates the confinement of the top 30% of the highest energy electrons locally within the nanometer-scale tube for the x-ray driven case. We see the accelerated electrons are broken into two main parts. This is because electrons are trapped nonconsecutively because of the nonlinear evolution of the wakefield. Moreover, electrons of energy below top 30% are excluded, so the "cut" phenomenon is much clearer. The transverse radius of the electron beam is almost 3 orders of magnitude smaller than that of the optical laser case, while the phase space remains nearly the same, which is beneficial to the beam emittance. According to the expression [33] for the beam emittance which is given by

$$\varepsilon_{N\_\mathrm{rms}} = \langle\gamma\beta\rangle\sqrt{\langle y^2\rangle\langle y'^2\rangle - \langle yy'\rangle^2},$$

where $\gamma\beta$ is the momentum, $y' = p_y/p_x$, and $p_x$, $p_y$ are the longitudinal and transverse momenta, the normalized emittance of the top 30% highest energy electrons for the x-ray case is about 0.0187 mm mrad, which is almost 3 orders of magnitude smaller than the 28.5 mm mrad for the optical laser case. Such an electron beam with low emittance and promising ultrahigh energy holds potential for the application for a future collider. In addition to this, there are also other promising advantages by using the nanotube, such as the better field structures for the significant improvement of the acceleration as concluded in Ref. [34] in the optical laser case. These may be observed in Figs. 2 and 3.

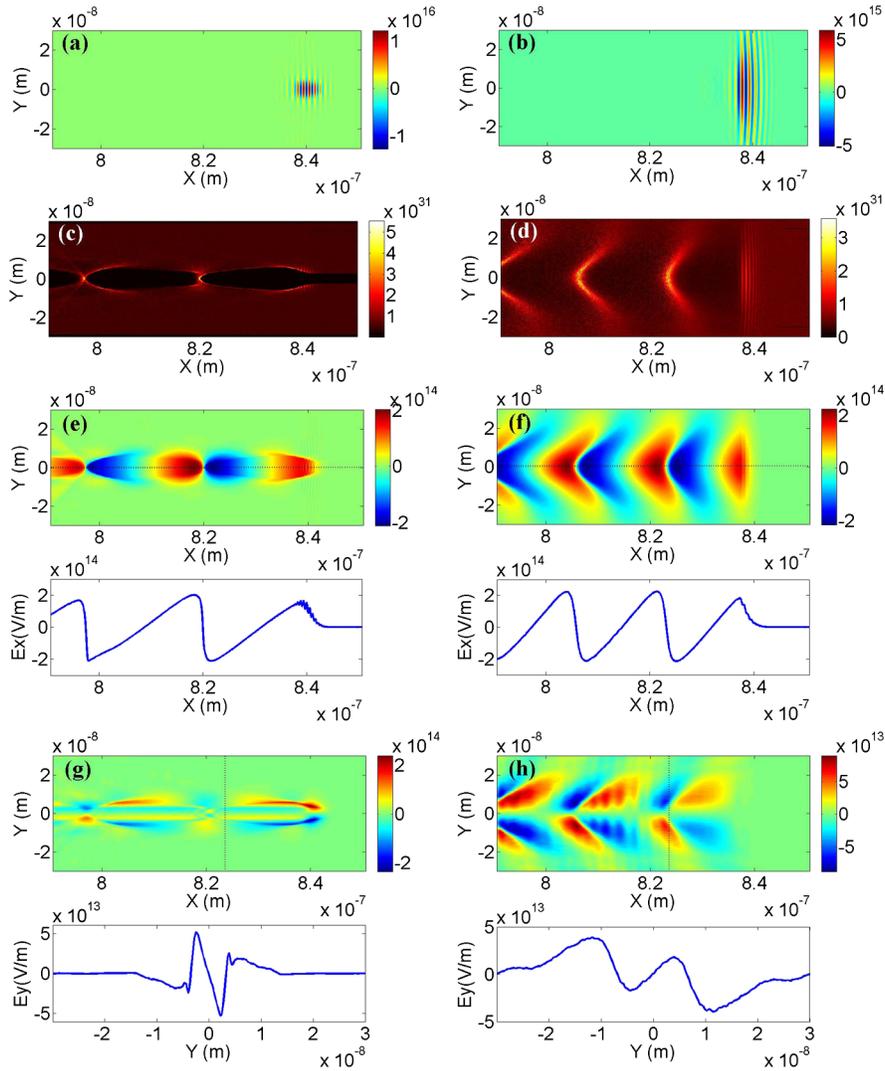

FIG. 2. The base case wakefield excitation with x-ray laser in a tube, in comparison with a wakefield in a uniform system. Distributions of (a) and (b) the laser field $E_z(V/m)$, (c) and (d) electron density $n_e(m^{-3})$, (e) and (f) longitudinal wakefield $E_x(V/m)$ including the $E_x$ lineout at $y = 0$ axis (the position of dot line), and (g) and (h) transverse wakefield $E_y(V/m)$ including the $E_y$ lineout at $x = 8.24 \times 10^{-7}$ m axis (the position of dot line) in terms of (a), (c), (e), and (g) tube and (b), (d), (f), and (h) uniform density cases driven by the x-ray pulse.

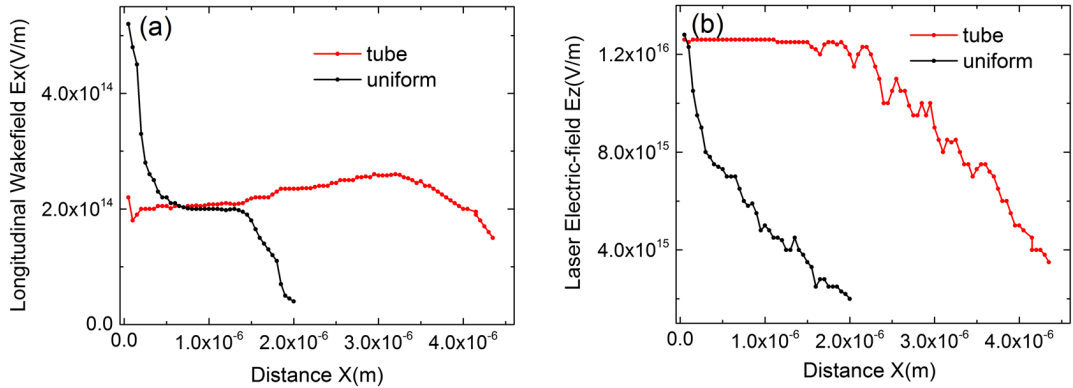

FIG. 3. Evolution of the maximum longitudinal wakefield $E_x(V/m)$ and the laser field $E_z(V/m)$ as the function of propagation distance $x(m)$ for both nanotube (red dotted line) and uniform plasma (black dotted line) cases with the same conditions as Fig. 2.

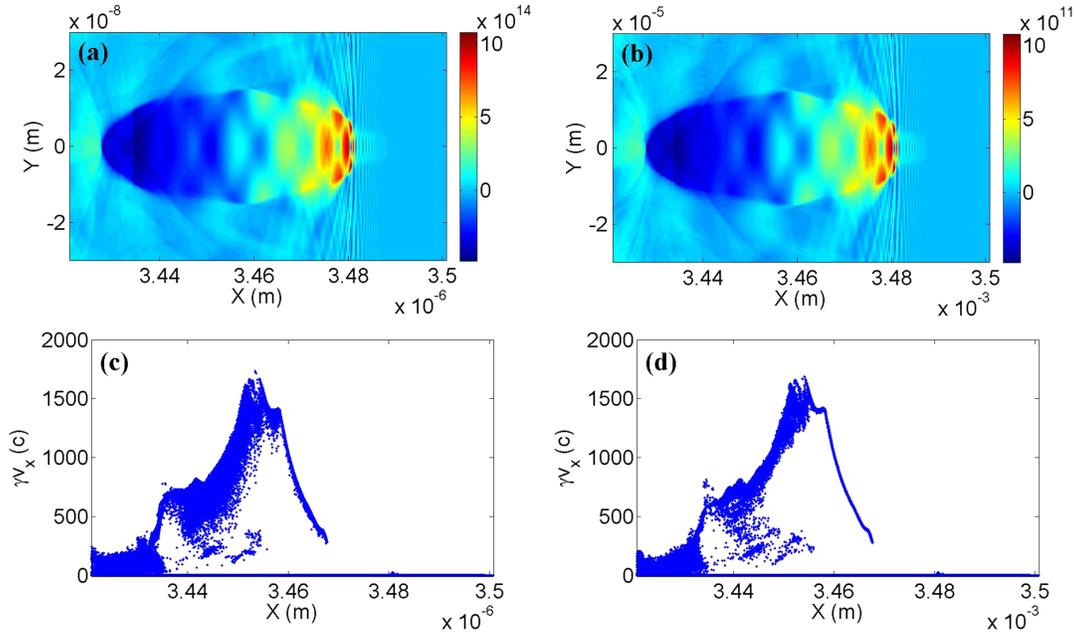

FIG. 4. Comparison and a certain scalability between the x-ray regime and the optical one. Distributions of (a) and (b) the longitudinal wakefield $E_x(V/m)$ and (c) and (d) electron longitudinal momentum $\gamma v_x$ induced by (a) and (c) the x-ray laser pulse and (b) and (d) 1 eV optical laser pulse in a tube when $a_0 = 10$.

In addition, the linear density of the top 30% and 80% highest energy electrons (that is, electrons of energy higher than $E_{e1} = 604$ MeV and $E_{e2} = 173$ MeV are considered) is $3.2 \times 10^{14}/m$ and $1.54 \times 10^{15}/m$ for the case of optical laser, and the linear density of the top 30% and 80% highest energy electrons (electrons of energy higher than $E_{e1} = 621$ MeV and $E_{e2} = 177$ MeV are considered) is $3.1 \times 10^{14}/m$ and $1.56 \times 10^{15}/m$ for the case of the x-ray laser. Here the top 80% is chosen instead of 100% to exclude the immobile background electrons. Assuming that the third dimension is proportional to the wavelength, the number of accelerated electrons in the x-ray laser case is $10^3$ times lower than that in the optical laser case. Here one point that should be emphasized is that the total laser energy $\varepsilon_L \sim a_0^2 \sigma_L^2 \sigma_x / \lambda_L^2$ which also scales with the laser wavelength, that is, the x-ray laser energy is $10^3$ times smaller than that of the optical laser. The energy transfer efficiency from laser to accelerated electrons is nearly unchanged. The emittance of the top 80% energetic electrons is 0.069 mm mrad for the x-ray laser, while the emittance of the top 80% energetic electrons is 64 mm mrad for the optical laser, $10^3$ times lower than that in the x-ray laser case.

TABLE II. Summary of the laser and plasma parameters for the electron acceleration case.

| Laser wavelength $\lambda_L$ | Normalized peak amplitude $a_0$ | Width radius $\sigma_L$ | Length radius $\sigma_x$ | Plasma density $n_{tube}$ | Tube radius $\sigma_{tube}$ |
|---|---|---|---|---|---|
| 1 nm | 10 | 5 nm | 5 nm | $5 \times 10^{24}$ W/cm$^3$ | 2.5 nm/0 nm |
| 1 $\mu$m | 10 | 5 $\mu$m | 5 $\mu$m | $5 \times 10^{18}$ W/cm$^3$ | 2.5 $\mu$m/0 $\mu$m |

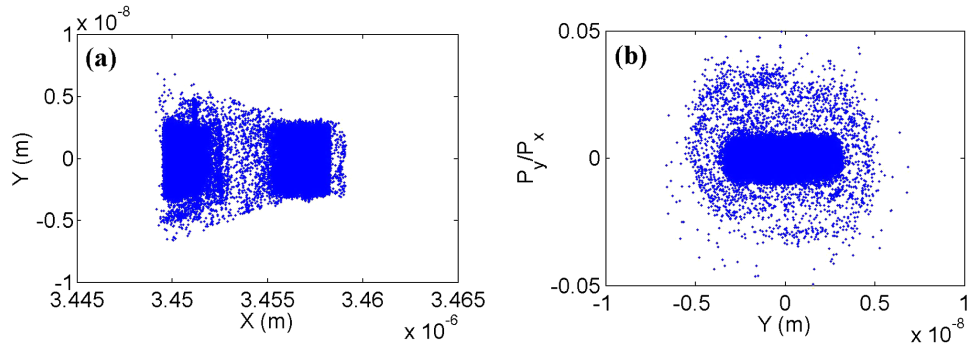

FIG. 5. (a) The accelerated electron beam quality in the x-ray wakefield in a tube. The space distribution $(x, y)$ and (b) the transverse phase space $(y, p_y/p_x)$ of the top 30% highest energy electrons in the case of x-ray laser. The parameters are the same as in Fig. 4.

It is well known that electrons also undergo betatron oscillations due to the transverse wakefield as they are gaining energy in the longitudinal wakefield and radiate x-ray or gamma-ray photons [35–38]. As mentioned above, when the ratio between the laser and the plasma wavelengths is constant, the energy gain is almost the same. However, the betatron radiation is quite different and cannot be normalized by the laser wavelength when QED effects are considered because the photon emission scales with the real electric field while the energy gain scales with the normalized laser amplitude $a_0$. As shown in Figs. 4 and 6, similar wakefield structure and energy gain, but quite different photon energy distributions in the two different laser wavelength cases are shown. Hundreds of keV to MeV photons are generated in the optical laser case, in which the high energy may be resulted from the wide oscillating radius since the injection position in the transverse direction depends on the tube diameter. On the other hand, although the electron energy gain is only several hundred MeV, which is almost the same as that of the optical laser case, the photon radiation energy is high to hundreds of MeV when undergoing the much stronger field in the x-ray laser case, which may be applied in astrophysics research, and cosmic ray generation.

According to the classical radiation theory [35,36], the photon critical energy due to betatron radiation scales with $\gamma^2 n_e r_\beta$, where $\gamma$ is the electron relativistic factor, $n_e$ is the electron density and $r_\beta$ is the betatron amplitude, which implies there should be an increase by a factor of $1/\lambda_L$ in the photon energy for the x-ray case over that of the optical laser case. However, the simulation including QED effects shows it to be smaller than this, or about the factor of 200 as seen in Fig. 6. This results from the quantum effects and can be partially explained by the replacement of $\nu \to \nu(1 + h\nu/E)$, which shows the quantum mechanical correction, where $\nu$ is the photon frequency and $E$ is the electron energy [39,40].

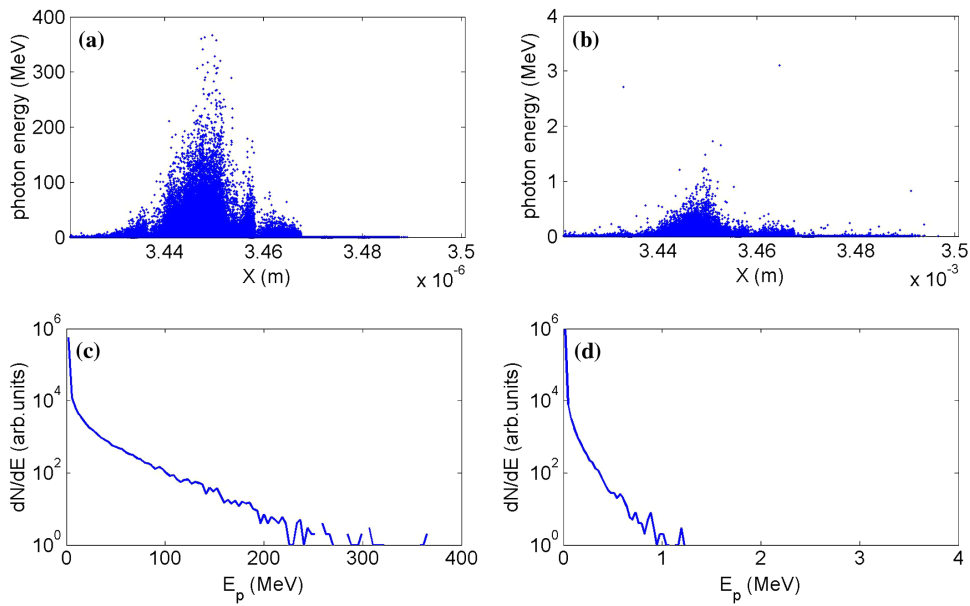

FIG. 6. The energy spectrum and spatial distribution of photon emitted from the wakefield driven by x-ray laser and optical one. (a) and (b) Photon energy distributions and (c) and (d) photon energy spectrum in the (a) and (c) x-ray driven case and (b) and (d) 1 eV optical laser driven case in a tube. The parameters are the same as in Fig. 5.

In addition, according to QED theory [41], if the QED parameter $\eta = \gamma E \sin\theta / E_{\text{crit}}$ is close to, or of order unity, the QED effects become important. Here $\theta$ is the angle between the electric field $E$ and the electron momentum, and $E_{\text{crit}} = 1.3 \times 10^{16}$ V/cm is the Schwinger field. At QED energies, electrons are expected to emit photons with a most probable value of $\hbar\omega_{mp} = 0.44\eta\gamma m_e c^2$. Using the maximum energy electrons in the wakefield $\gamma_{\max} = 1700$, which experiences a transverse wakefield of $E \sim 5 \times 10^{12}$ V/cm [of the order of longitudinal wakefield as shown in Fig. 4(a)], the maximum QED parameter is found to be $\eta_{\max} = 0.577$. Therefore, the expected value of most probable photon energy corresponding to the highest energy electrons is predicted to be $\hbar\omega_{mp} = 221$ MeV. This expected value agrees reasonably well with the observed photon cutoff value $E_{p-\text{cutoff}} = 300$ MeV as shown in Fig. 6(c).

Here one point should be noted that there is nearly no difference of the electron energy between Figs. 4(c) and 4(d), which means there is little radiation reaction effect in the x-ray laser case although the photon energy is much higher than that in the optical laser case. By comparing with the case in which the radiation reaction was turned off for the x-ray laser case, the results turn out that little radiation reaction effect on electron beam dynamics, i.e., energy and emittance are found. According to the Landau-Lifshitz prescription [41], the ratio of the damping force (radiation reaction force) to the ordinary Lorentz force scales with $\gamma^2 E$. In the x-ray laser case, $E$ is about $10^3$ times higher, which means the radiation reaction force becomes much more important. However, the energy loss of electrons (radiation reaction effect) depends on the acting time/distance, so radiation reaction effect in the x-ray laser case is possible to be weak because of the much shorter ($10^3$ times) acting time/distance. In addition, the ratio of the total energy of all photons to the total energy of all electrons is $9.4 \times 10^{-3}$ in the x-ray laser case. That means the radiation reaction effect is still weak, although the value is much higher than that of $2.2 \times 10^{-5}$ in the optical laser case.

According to the above analysis, the effect of the laser frequency on the betatron radiation is expected to be quite important under the condition of the same laser power. For the current laser level in the near term, the triple frequency ($3\omega$) laser can be chosen to drive a more intense wakefield and get higher energy photons. Simulations have confirmed that, driven by the triple frequency laser, the obtained average photon energy is at least twice that from the fundamental one, which can be considered as an effective approach to increase the radiation photon energy.

## III. THE SCALINGS IN THE X-RAY REGIME AND DISCUSSION

Here we survey the property of x-ray wakefield acceleration in nanotubes with respect to several parametric scalings. Compared with the uniform density case, the laser pulse can be well confined in the nanotube and propagate over a longer distance. Therefore, the tube radius is critical to the wakefield and the acceleration in addition to the other two common parameters, i.e., laser intensity and tube density. Figure 7(a) shows the result when the tube radius varies while the other parameters are kept the same. The wakefield begins high due to the nonlinear evolution when the tube radius is small. As the radius ratio goes up, the effective density decreases. This results in a decrease in wakefield strength since it is proportional to the density, and energy gain is expected to increase if the ratio is not too large because the acceleration length is extended. In the present case, the wakefield scales with the tube radius ratio as $E_x \propto (\sigma_{\text{tube}}/\sigma_L)^{-1.827}$. It should be noted that when the tube radius is small or can be compared to the laser pulse width, such as the present cases, the physics becomes closer to a uniform plasma case. On the other hand, for much wider tube cases, the wakefield becomes less intense and deviates away from the uniform plasma wakefield acceleration. For this case, the physics may more closely resemble the dielectric wakefield acceleration [33,42–44]. However, the driving unwanted higher order mode (dipole, quadrupole, etc.) in this case may be an issue, which is different with the present plasma regime where only the plasma frequency is important for acceleration. When the channel radius is fixed, for example as in Fig. 7(b), $\sigma_{\text{tube}}/\sigma_L = 1$, at first the wakefield increases along with the tube density ($n_{\text{tube}}$) but tends to saturate because it is hard to excite a wakefield when the parameters of the laser pulse and density are mismatched. More importantly, the significant feature is that lower density results in higher energy gain. The wakefield scales with the tube wall density as $E_x \propto n_{\text{tube}}^{0.47}$ in the low density region. When the laser intensity ($a_0$) goes up, the effective density grows higher because of the increasing plasma wavelength. In the present case, the wakefield scales with the laser field as $E_x \propto a_0^{1.875}$ when $\sigma_{\text{tube}}/\sigma_L = 0.5$ and $E_x \propto a_0^{1.763}$ when $\sigma_{\text{tube}}/\sigma_L = 1$, which shows a similar scaling for different radius ratios.

Compared with the linear wakefield theory $E_x \propto (\omega_p) \sim n_e^{1/2}$, 1D nonlinear theory $E_x \propto (a_0^2/2)(1 + a_0^2/2)^{-1/2}\omega_p$ [9] and 3D nonlinear (Bubble) theory $E_x \propto a_0^{1/2}\omega_p$, $\omega_p \propto n_e^{1/2}a_0^{1/2}$ [8] in the uniform plasma, as well as the previous theoretical results on the x-ray wakefield accelerator in solid-density plasma channels $E_x \propto \omega_p P^{1/2} \sim n_e^{1/2} a_0$ [45–47], the wakefield scaling with the wall density in our nanotube case in principle agrees with the theory expected as $E_x \propto n^{1/2}$ in the uniform density case and the wakefield scaling with the laser intensity is close to the 1D nonlinear theory.

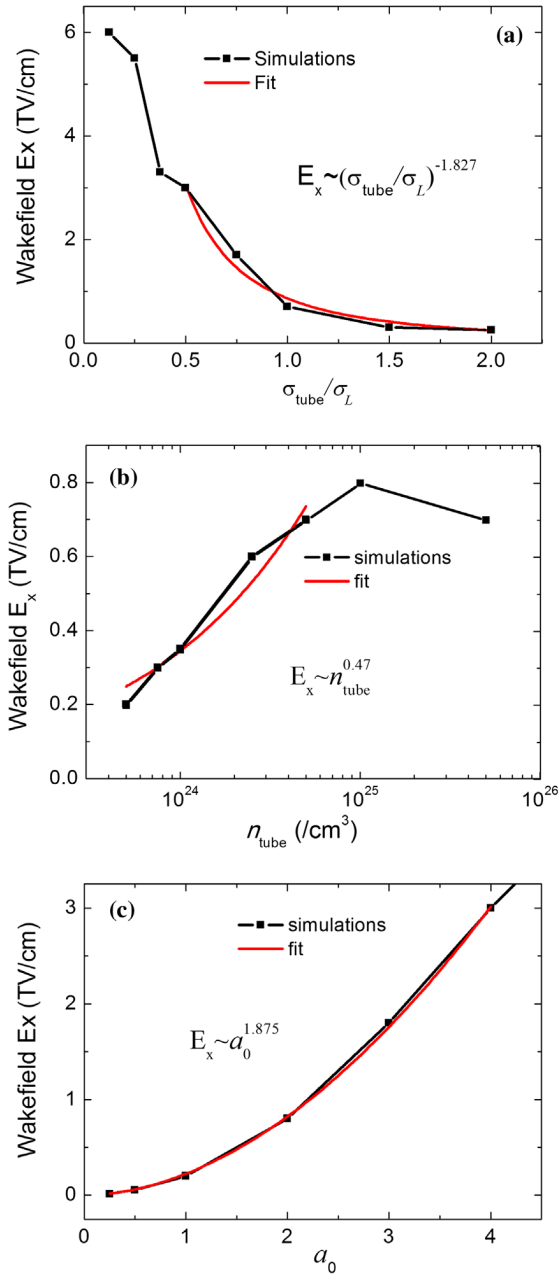

FIG. 7. Wakefield scalings in the x-ray regime with (a) tube radius, (b) tube wall density when the tube radius is fixed $\sigma_{\text{tube}}/\sigma_L = 1$, and (c) laser intensity when the tube radius is fixed $\sigma_{\text{tube}}/\sigma_L = 0.5$.

For the parameters in the above discussion, there are several technological challenges to be considered in the experiments, such as the x-ray intensity and wakefield strength, the focus of such x-ray beam onto a nanometer size tube, and the compression of the present state-of-the-art coherent x-ray pulse down to a few nanometers. In reality, in the short term the parameters can be extended to the acceptable region since the results scale with the laser wavelength.

On the other hand, as we know, wakefields may be created not only by lasers, but also by a beam of electrons or ions because the plasma responses to these drivers are essentially the same [2,48]. Beam driven acceleration in ultradense plasma, including a hollow plasma channel with density of $10^{25}$–$10^{28}/\text{m}^3$ has been explored [10,11] and a high acceleration gradient of TeV/m has been obtained [22]. Moreover, a hollow channel, as a more efficient structure in controlling beam parameters in the dense plasma interaction, has been confirmed. In this beam driven case, beam density is especially critical to increasing the acceleration gradient, just as with the laser intensity. When the beam density is high enough to be compared with that of the dense plasma, such as $10^{30}/\text{m}^3$, and the beam size is small enough to be at nanometer scale (however, these are tall orders in the current beam technology), similar results with that driven by the x ray are expected. In the short term, particle beams instead of x-ray lasers can be used as the driver in a nanotube. For current particle beams, as they pass through a nanomaterial, a periodic pattern such as wakefield is expected to be generated. Such patterns can serve as optical elements for the beam, and, correspondingly, linear phenomena such as diffraction, beam bending or focusing are expected to be exhibited.

## IV. CONCLUSION

In conclusion, owing to the latest invention of the thin film compression technique, one single-cycled optical laser pulse can in principle be converted into a coherent ultrashort x-ray pulse via relativistic compression. A new and promising scheme employing such an x-ray driven wakefield in a nanotube has been demonstrated by a computer simulation for a compact accelerator to attain an ultrahigh acceleration gradient for charged particle acceleration. In this case, an acceleration gradient of TeV/cm is generated and high energy electrons with much lower emittance are obtained in such a wakefield. In a very narrow limit of the tube, the energy scalings resemble those in the uniform medium formulation. In addition to the aspect of acceleration, under the x-ray driven nanotube wakefield scheme, hard photons with energies at $\sim O(100 \text{ MeV})$ are emitted. Those may be invoked as a tool to serve as a novel light source in very high energies in a compact fashion and to explore more unknown physics, although there are several technological challenges in the future in the realization of the experimental operation and parameters suggested in our work. These include such an x-ray laser pulse generation and the manipulation of such small size laser and target. In this regard the recent thrust in ultraintense laser developments leads us to a high hope that such projects can accelerate the progress in this new exciting field with an added impetus.

## ACKNOWLEDGMENTS

The work has been supported by the Norman Rostorker Fund, and was further supported by the National Natural Science Foundation of China (Grants No. 11374319, No. 11125526, No. 11335013, No. 11674339 and No. 11127901), the Ministry of Science and Technology of the People's Republic of China (Grant No. 2016YFA0401102), the Strategic Priority Research Program of the Chinese Academy of Sciences (Grant No. XDB16), and the China Scholarship Council. We thank the dedicated effort of Extreme Light Infrastructure–Nuclear Physics in realization of the thin film compression [49]. Toshiki Tajima thanks the Einstein Professorship of the Chinese Academy of Science and Professor Ruxin Li for his kindness to have hosted Toshiki Tajima as Einstein Professor, from which this work started and which in part supported this work, including that of Xiaomei Zhang.